# Ancient and recent collisions revealed by phosphate minerals in the Chelyabinsk meteorite

Craig R. Walton[1], Oliver Shorttle[1,2], Sen Hu[3], Auriol S. P. Rae[1], Ji Jianglong[3], Ana Černok [4], Helen

Williams[1], Yu Liu[5], Guoqiang Tang[5], Qiuli Li[5], & Mahesh Anand[4]

1 - Department of Earth Sciences, University of Cambridge, Downing Street, Cambridge CB2 3EQ,

UK

2 - Institute of Astronomy, University of Cambridge, Madingley Road, Cambridge, CB3 OHA, UK

3 - Key Laboratory of Earth and Planetary Physics, Institute of Geology and Geophysics, Chinese

Academy of Sciences, Beijing, 100029 China

4 - School of Physical Sciences, Open University, Walton Hall, Milton, Keynes, MK7 6AA, United

Kingdom

5- State Key Laboratory of Lithospheric Evolution, Institute of Geology and Geophysics, Chinese

Academy of Sciences, Beijing, 100029 China

**The collision history of asteroids is an important archive of inner Solar System evolution. Evidence for these collisions is brought to Earth by meteorites, which can preserve impact-reset radioisotope mineral ages. However, as meteorites often preserve numerous mineral ages, their interpretation is controversial. Here, we combine analysis of phosphate U-Pb ages and**

**mineral microtextures to construct a collision history for the highly shocked Chelyabinsk meteorite. We show that phosphate U-Pb ages in the meteorite are independent of thermal history at macro-to-microscales, correlating instead with phosphate microtexture. Isotopic data from pristine phosphate domains is largely concordant, whereas fracture-damaged domains universally display discordance. Combining both populations best constrains upper (4,473 ± 11 Ma) and lower intercept (-9 ± 55 Ma, i.e., within error of the present day) U-Pb ages for Chelyabinsk phosphates. We conclude that all phosphate U-Pb ages were completely reset during an ancient high energy collision. Fracture-damaged phosphate domains experienced further Pb-loss during mild collisional heating in the geologically recent past, and must be targeted to properly constrain a lower intercept age. Targeting textural sub-populations of phosphate grains can significantly improve the calculation and interpretation of U-Pb ages, permitting more robust reconstruction of both ancient and recent asteroidal collision histories.**

Collisions play a fundamental role in shaping rocky objects in our Solar System by (i) building protoplanets [1], (ii) replenishing or eroding planetary atmospheres [2], and (iii) violently perturbing surface environments [3]. Collisions shape surfaces through cratering; transforming the mineralogy and texture of affected rocks by shock metamorphism, and driving diffusive resetting of radioisotope mineral ages through post-shock thermal metamorphism [4]. Crustal records of impact bombardment on the Earth, Moon, and Mars have been used to validate dynamical models of Solar System evolution [5, 6], but suffer from the effects of planetary resurfacing in deeper time.

Asteroids provide an alternative record of collisional events in the inner Solar System [7]. Unlike planets, asteroids have been thermally quiescent (cold) since around 4500 Ma [8]. Therefore, any mineral ages younger than the end of parent body metamorphism should faithfully record impact-induced metamorphism. Phosphate minerals represent a wide-spread class of low- to-medium closure temperature U-Pb geochronometers found in meteorites. Age clusters in the meteorite phosphate U-Pb record have been linked to the formation of Earth's Moon [7, 9], the migration of giant planets [10], and the recent to long-term evolution of the asteroid belt [11, 12, 13, 14, 15, 16]. Constraints on key events in Solar System and Earth history are therefore written in the collision histories of meteorites. However, this seemingly ideal data set is compromised by ambiguity in the interpretation of upper versus lower concordia intercept phosphate U-Pb ages.

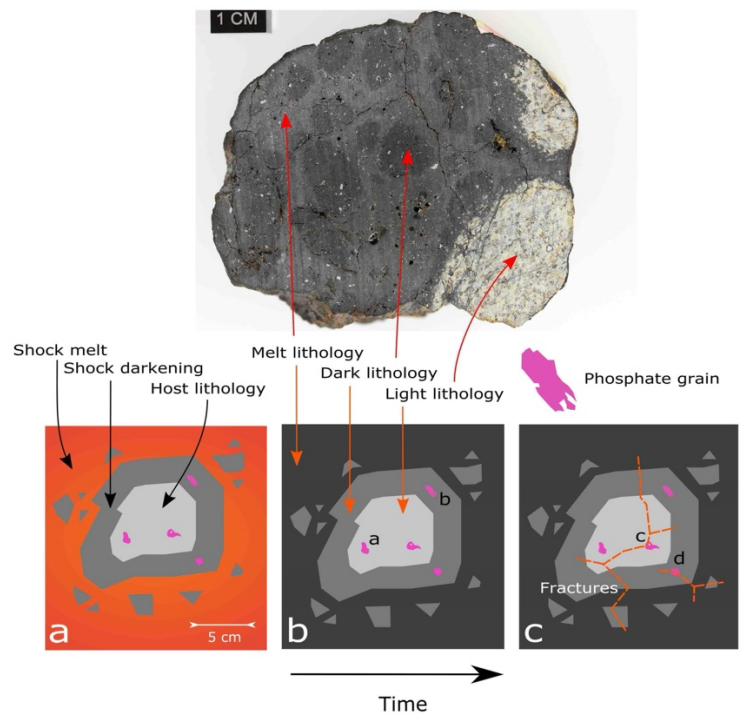

**Figure 1** Rock textures in Chelyabinsk. Panels a-c: evolution of Chelyabinsk breccia, from (a) initial formation during shock-melting, brecciation, and shock darkening of host rock material, through (b) solidification into light, dark, and melt lithologies, and (c) subsequent minor disturbances, such as the propagation of fracture networks. Pink symbols represent host-rock phosphate minerals, which are only found in the light and dark lithologies. Photograph used is of Chelyabinsk specimen NHMV-O707; Credit Ludovic Ferriere-NHM Vienna, Austria.`

This ambiguity is exemplified by Chelyabinsk: an ordinary chondrite shocked meteorite sampling the LL asteroid, not affected by terrestrial alteration [17] (Figure 1). Chelyabinsk represents an allochthonous (formed from mobilised material), proximal impactite (short transport distance from point of impact), clast-rich (containing pieces of host-rock material) melt rock, sometimes known as impact melt breccia [18]. Melt rocks are formed during high velocity collisions, which deliver sufficient energy to induce extensive melting of the target object [19]. Chelyabinsk preserves three lithologies: light (host rock), dark (containing a higher proportion of melted phases), and shock-melt (fully melted and quench crystallised material) (Figure 1). Phosphates in the dark lithology experienced peak temperatures at least 200 K higher than those in the light lithology, whilst phosphates in the melt lithology were destroyed [20, 21, 22] (Figure 1).

The simplest interpretation of these observations is that all three lithologies were produced together during a single impact event: light lithology fragments were entrained in shock-melt and the dark lithology formed by interaction between the two, as both individual isolated blocks and as cooked margins around larger light lithology fragments [23, 17, 24] (Figure 1). However, both previously reported upper (4,456 ± 18 Ma) and lower (559 ± 180 Ma) intercept phosphate UPb ages for Chelyabinsk [25, 26] have been individually suggested to record the same high-energy event of simultaneous melting and brecciation. Furthermore, numerous other ages are obtained using different mineral chronometers [25, 27, 28, 23, 29, 30] (Supplementary Figure 18). This level of ambiguity in the collisional chronology of shocked meteorites draws a veil over key events in Solar System history, which could otherwise be constrained using phosphate mineral ages. To resolve this apparent paradox, we require a better understanding of the phosphate texture-age record of asteroids.

Mineral microtextures provide geological context for spatially resolved radioisotope ages, e.g., crystal structural integrity, which can influence Pb diffusion [31]. Microtextures are increasingly being targeted to reduce uncertainty in the interpretation of spatially resolved phosphate U-Pb ages [32, 33, 34, 35, 36]. We have previously conducted a detailed microtextural survey of phosphate minerals in the Chelyabinsk meteorite [20]. Here, we present an in-situ U-Pb dating study of texturally-distinct phosphate populations in the meteorite, allowing us to re-interpret the collision history of Chelyabinsk and its parent body. We utilise Scanning Electron Microscopy (SEM), Electron Back Scattered Diffraction (EBSD), Cathodoluminescence (CL), and Secondary Ionisation Mass Spectrometry (SIMS) analyses to assess the phosphate texture-age record of Chelyabinsk (see Methods). We further verify our interpretative model by making and testing predictions for the wider meteoritic phosphate texture-age record.

Results and Discussion

*Microtextural evidence for recent and ancient collisions*

The light and dark lithologies of Chelyabinsk each preserve grains of the phosphate minerals apatite ($Ca_5(PO_4)_3[OH,Cl,F]$) and merrillite ($Ca_9NaMg(PO_4)_7$). Previous EBSD analyses [20] have revealed that all light lithology phosphate grains display domains of distorted crystal orientation, which is most likely attained during a crystal-plastic recovery process (Figure 2a), whereas dark lithology merrillites display randomly oriented strain-free sub-domains (Type II recrystallisation) [37], evidencing recrystallisation likely driven by cataclastic deformation during more intensive heating (Figure 2b). This specific association of phosphate textures with lithology type suggests the formation of the Chelyabinsk melt rock during a *single impact event* - an interpretation which is supported by similar observations made for phosphates in terrestrial impactites [24].

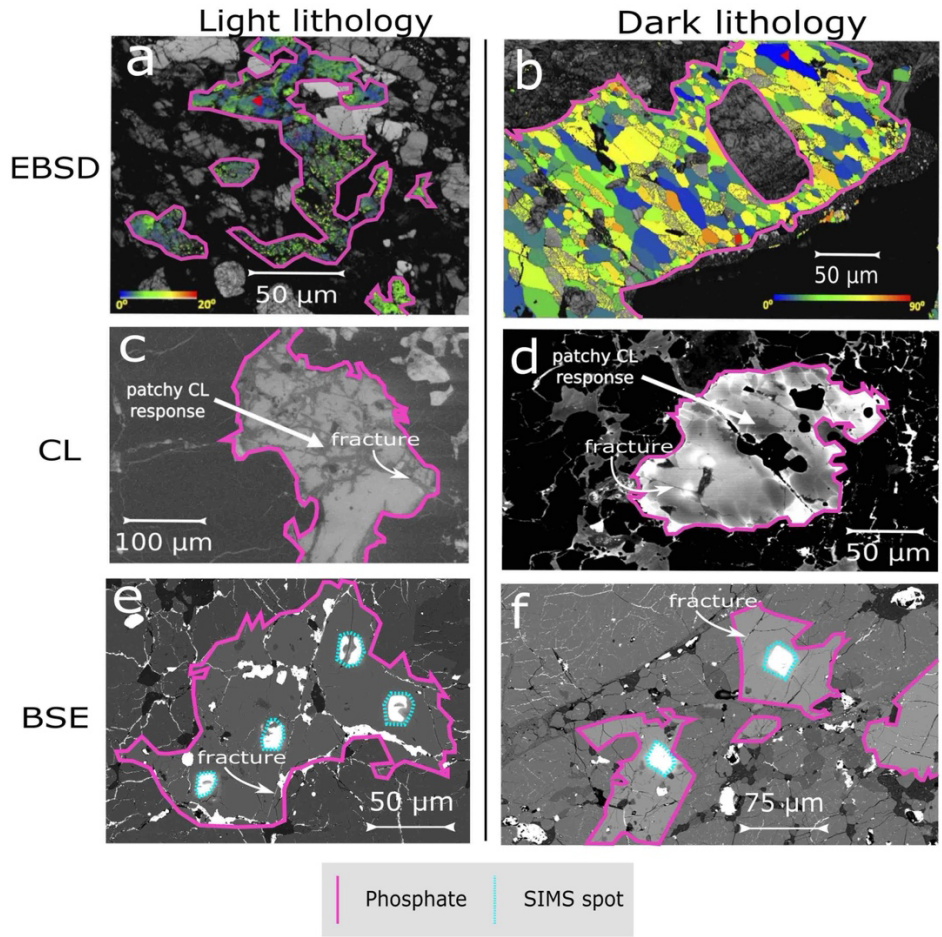

**Figure 2** Mineral microtextures in Chelyabinsk. In EBSD images, color-scheme indicates crystal lattice misorientation relative to an arbitrary point (red triangle). a) apatite grain showing smooth gradations in lattice misorientation, revealing strain and associated deformation. This strain was most plausibly accumulated during impact. b) merrillite grain with distinct subgrains of uniform and unstrained crystal lattice orientation, revealing recrystallisation that likely developed in response to more extensive heating. c) apatite showing patchy CL response, correlated with fractures. d) merrillite showing subgrain recrystallisation, as well as overprinting patchy CL response correlated with fractures. e) apatite showing extensive fracturing. Metal and sulfide

veins (white in BSE image) fill some fractures, whereas others are unfilled. f) apatite grain showing similar fracturing, proximal to a shock-melt-vein. Partially annealed metal and sulfide veins are abundant in the silicate matrix. In all images, phosphates are outlined in purple. Pole figures and further data related to panels a and b are available in Supplementary Information section 1.

The strain-free domains of dark lithology merrillite indicate minimal post recrystallisation deformation. However, CL images, which are sensitive to phosphate trace element composition [33], reveal patchy textures correlated with fractures, which are clearly visible in Back Scatter Electron (BSE) images (Figure 2c-f). These features evidence a later low energy event, which affected individual grains in both the light and dark lithology, regardless of their microtextural state (Figure 2c-d). Phosphate microtextural evidence therefore records distinct high temperature pathways in the dark and light lithologies during primary impact, whilst shared patchy CL textures (Figure 2cd) and fracture networks (Figure 2e-f) indicate equivalent later (minor) shock histories (Walton et al., 2021).

*Possible scenarios of age resetting*

Lead diffusion in apatite is strongly temperature dependent [38]. Complete diffusion and loss of Pb will also occur faster for smaller crystals. The deformed and recrystallised populations of phosphates in Chelyabinsk experienced different thermal histories (Figure 1), and have grain sizes (maximum width) that vary from sub-micron (below minimum detection size by EBSD) to several hundred micrometres throughout each lithology (Figure 2). These observations allow us to test different hypotheses regarding the nature and timing of collisions recorded by Chelyabinsk (Figure 3).

If recent shock and post-shock thermal metamorphism was responsible for simultaneously inducing strain, recrystallisation, and U-Pb discordance in Chelyabinsk phosphates, we predict that Pb-loss should be significantly more extensive in the more intensely shocked and heated dark lithology phosphates [39, 37], as well as in smaller grains (Figure 3b). Conversely, late Pb-loss may be unrelated to the development of Chelyabinsk's primary shock textures, and may instead record a milder recent thermal event. In this case, we predict that Pb-loss should be uncorrelated with phosphate location in Chelyabinsk light or dark lithology (Figure 3c). This holds because (1) lithology-scale temperature-dependent Pb loss would have occurred in young minerals, with less accumulated Pb as compared to the present; (2) to the extent that phosphate populations were only partially reset during an initial energetic collision, billions of years of subsequent radiogenic decay will have greatly reduced space on the concordia plot between grain populations that experienced Pb-loss early; and (3) reverse discordance induced in late events may have further blurred the original data distributions of light versus dark lithology phosphates.

For comparison, mild temperatures which are insufficient to drive Pb-diffusion from pristine zicon ($ZrSiO_4$) crystal lattice may be sufficient to drive Pb-loss from damaged zircon grains [40, 31]. Much as in metamict zircons that do not experience annealing, then, phosphate crystal domains damaged by fractures may be susceptible to a late episode of Pb-loss at only mild temperature conditions. In this scenario, we predict that damaged crystal domains will display more extensive Pb-loss, or discordance, whilst pristine domains will be broadly concordant (Figure 3c). These hypotheses allow us to use textural and chronometric relationships to constrain the timing and nature of collisional events affecting the Chelyabinsk melt breccia.

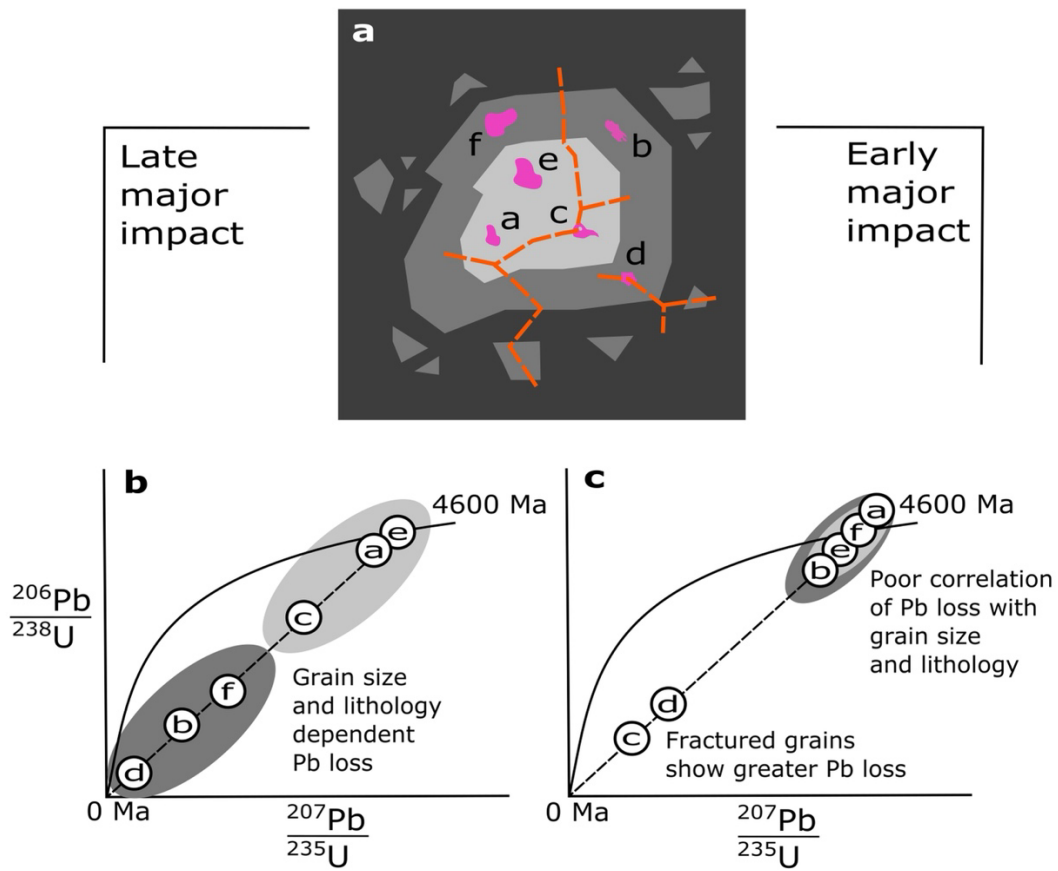

**Figure 3 Scenarios for Pb-loss from Chelyabinsk phosphates during impact.** a) Schematic view of phosphate grain populations in Chelyabinsk lithologies. Small and large pristine phosphate grains are shown, as well as grains with fracture-damaged domains. b) If recent Pb-

loss corresponds to a major event of shock metamorphism and post-shock heating, Pb should diffuse more rapidly from smaller grains, and from those grains hosted in the dark lithology (higher peak temperature). c) If recent Pb-loss corresponds to a relatively mild heating event that only noticeably affected fractured-damaged grains, Pb-loss will have been stochastic, possibly independent of grain size, occurred throughout both lithologies equally, and will be more developed in fracture-damaged rather than pristine grain populations.

*Statistical analysis of phosphate U-Pb data*

Our results allow us to test the scenarios presented in Figure 3 for late Pb-loss in response to either (1) a primary impact event, with Pb-loss principally occurring in dark lithology grains that experienced more intensive heating, or (2) a mild secondary event, with Pb-loss principally occurring in damaged grains in both lithologies. Regressions of U-Pb data split by lithology and by microtextural state are shown in Figure 4.

We find no correlation of phosphate discordance with U-content or grain size in Chelyabinsk (Supplementary Figures 10-15). We tested the null hypothesis that the there is no difference in Pb-loss between each phosphate population using Two-sided Kolmogorov–Smirnov (KS) tests of 207-Pb/235-U data distributions. Light and dark lithology grain populations are statistically identical in this test (Figure 4c), whereas pristine and fracture-damaged populations are highly significantly different (Figure 4d). Our results support a scenario in which Pb-loss occurred mainly in fracture-damaged grains as a result of a late impact event, which only produced mild pressures and temperatures in the Chelyabinsk material.

There is then the question of how best to regress phosphate population data and interpret the resulting concordia intercept ages (Figure 4). At the 2 $\sigma$ level, upper and lower intercept age uncertainties obtained using light and dark lithology or pristine and fracture-damaged phosphate populations overlap (Figure 4). Pristine domains yield a well defined upper intercept age (4,453 ± 36 Ma) as well as a weakly constrained lower intercept age (696 ± 813 Ma). Fracture-damaged phosphate crystal domains yield a similarly well-constrained upper intercept age (4,477 ± 12 Ma) and a much more tightly constrained lower intercept age (-3 ± 56 Ma, i.e., recent, within error of the present day). We used F-tests to test the null hypothesis that all data should be regressed together, rather than being regressed as sub-populations. Results reveal that treating light and dark and pristine and fractured phosphate populations separately during regression is not statistically justified at 99 % confidence (Table S1). Pristine and fracture-damaged grains therefore serve to constrain different regions along a single linear regression, together yielding our preferred intercept ages of 4,473 ± 11 Ma and -9 ± 55 Ma. This upper intercept age is statistically identical to those previously reported for Chelyabinsk, whilst our revised lower intercept is several hundred Myr younger than previously reported ages [26, 25].

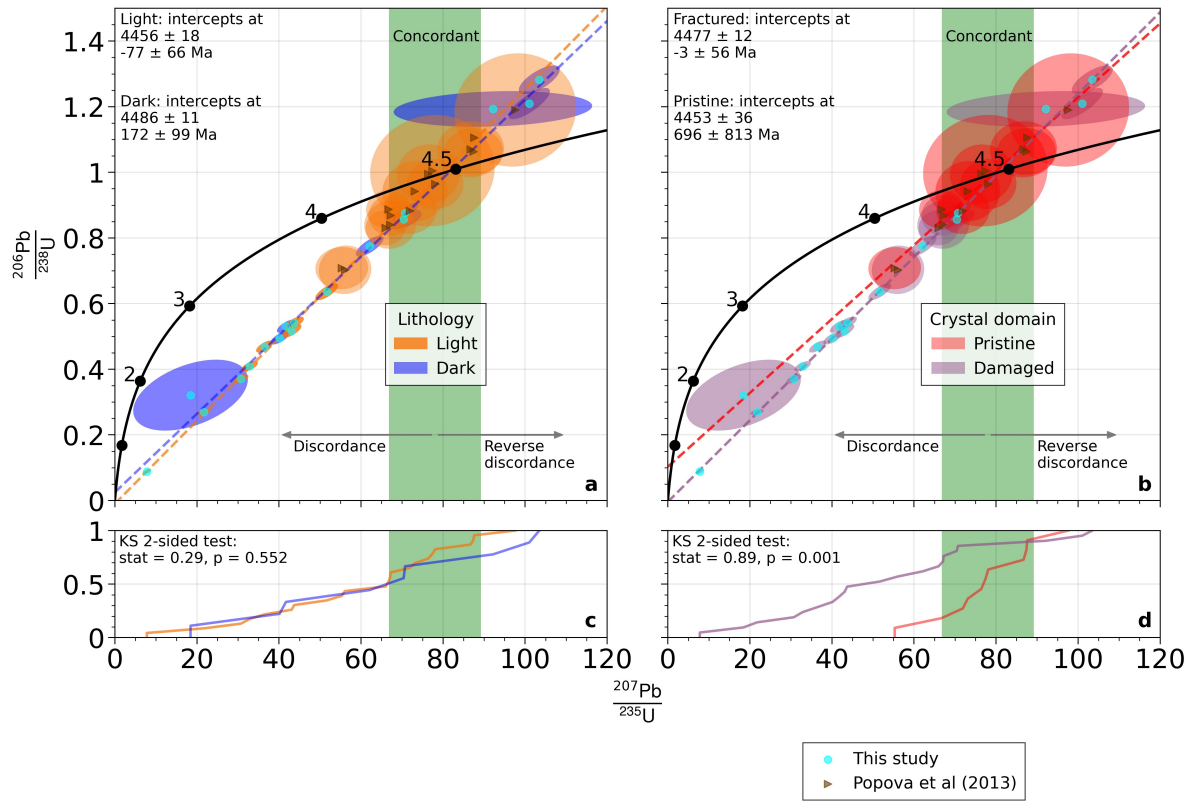

**Figure 4 Statistics and concordia chronology of phosphates in Chelyabinsk.** Each data point is shown with shaded 1$\sigma$ error ellipse. a) Comparison of light and dark lithology phosphate grain populations. b) Comparison of pristine and damaged phosphate crystal domain analyses. c) Statistical comparison of Cumulative Distribution Function (CDF) plots of 207-Pb/235-U ratios for light and dark lithology phosphate data. The populations cannot be statistically resolved using two-sided Kolmogorov–Smirnov tests. d) Statistical comparison of CDFs for pristine and damaged phosphate crystal domain data. The populations are statistically resolved, with the pristine population having higher Pb/U than the damaged grain population.

*Interpretation of U-Pb regressions and intercept ages*

The revised lower intercept obtained after identifying and including damaged phosphate domains in a phosphate U-Pb age calculation for Chelyabinsk appears to have geological significance. Lower intercepts may be of dubious meaning when no concordant data is observed [31]. However, many pristine phosphate domains display fully concordant spot data (Figure 4b). Multiple episodes of partial Pb-loss from damaged grains, which would greatly complicate any interpretation, should manifest as U-Pb spots that fall off the regression line [31]. However, isotope data for damaged crystal domains are well described by a single linear regression (Figure 4b). We conclude that the youngest and most tightly constrained U-Pb lower intercept age defined by fracture-damaged phosphates in Chelyabinsk most plausibly reflects Pb loss from damaged grains during a comparatively minor shock and reheating event in the geologically recent past [40, 31] (Figure 3c).

The large uncertainty on the lower intercept age obtained using pristine phosphate domains alone (Figure 4b) suggests that fracture-damaged phosphate grains must be identified and used in the regression in order to properly constrain lower intercept ages. Given that the preferred Chelyabinsk upper intercept U-Pb age of all phosphate domains presented here (4,473 ± 11 Ma) is younger than the time that primitive asteroids cooled below the Pb diffusion closure temperature for phosphate minerals (Figure 5), and given the similar degree of partial Pb loss from phosphates in both lithologies, all phosphate U-Pb ages must initially have been fully reset during a primary impact event (Figure 5). Partial Pb loss must then have occurred much later (Figure 5), following regeneration of Pb by U-decay (Figure 3c).

These data support a scenario in which an early energetic primary collision produced the light-dark textured melt breccia material, deforming, recrystallising, and damaging phosphates (Figure 5). A late second collision then liberated the Chelyabinsk breccia as spall, low-velocity ejecta, or catastrophic fragmentation of the parent body (Figure 5), subjecting the material to mild pressure-temperature conditions, further propagating fracture networks, and inducing Pb-loss from damaged phosphate grains. Our interpretation is consistent with evidence for enhanced Pb-loss from previously impact-metamorphosed phosphate grains [36], and from U-series disequilibria for recent impact-induced Pb mobility at mild temperatures in carbonaceous chondrites [41]. Our results support an emerging dichotomy between mechanisms of meteoritic and terrestrial apatite Pb-loss [37], with microtextures efficiently driving Pb-loss from extra-terrestrial phases that largely lack common Pb [36].

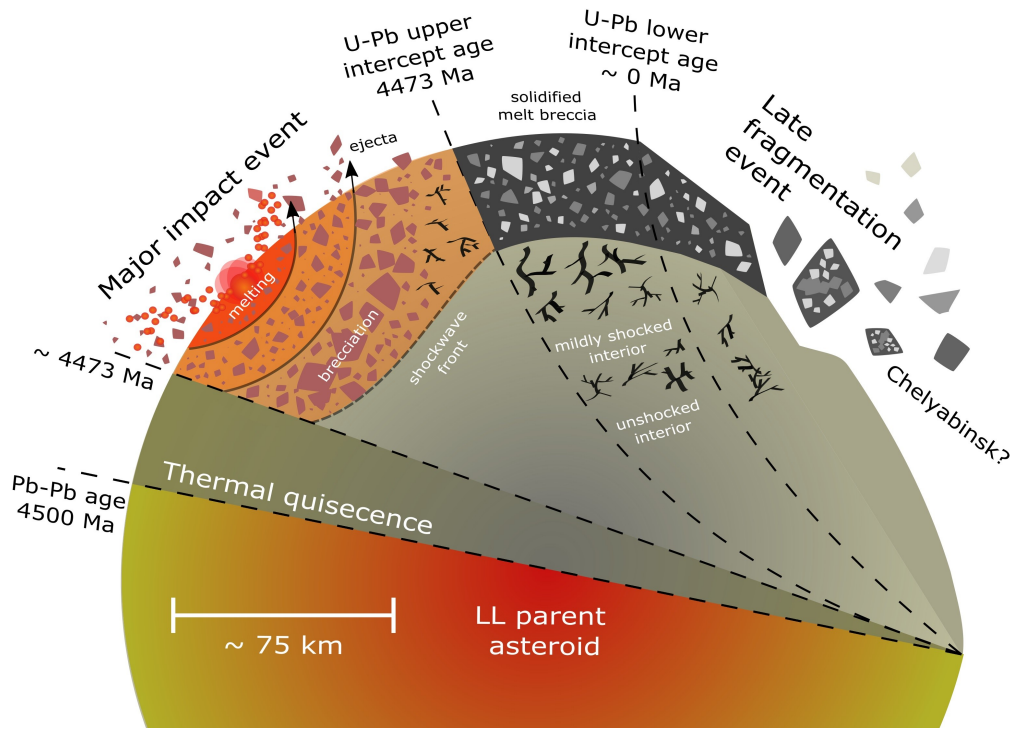

**Figure 5 Proposed formation history of the Chelyabinsk meteorite** a) Schematic view of collisions affecting the source material of Chelyabinsk on the LL parent body. Early radiogenic metamorphism gives way to cooling of the parent body. A primary collision locally produces shock-melt at the surface of the asteroid, which cools relatively quickly (shock-induced features incompletely annealed during post-shock thermal metamorphism). A later spalling event is then needed to liberate the Chelyabinsk material from the LL parent body, and may be responsible for late Pb loss and lower intercept ages.

*Structure of the meteoritic phosphate U-Pb record*

We can further test our model for the collision history of Chelyabinsk by using it to make predictions for the wider chondritic phosphate texture-age record. We group meteorites into highly shocked (S4-6) and weakly shocked (S1-3), which corresponds to the conditions above and below the threshold for phosphate U-Pb resetting determined by Blackburn et al (2017) [38]. If generally applicable, our model predicts that highly shocked meteorites should have fully reset upper intercept phosphate U-Pb ages (i.e., ages younger than the parent body cooling age of circa 4500 Ma), whereas phosphates in weakly shocked meteorites will record parent body cooling (ages greater than 4500 Ma). Both highly and weakly shocked meteorites may, but do not have to, display well-defined lower intercept ages, plausibly corresponding to a recent collision experienced by an asteroid.

Compiling all published SIMS single phosphate U-Pb ages for chondritic meteorites (Figure 6), we find support for our predictions. Our preferred upper intercept age of 4,453 ± 36 Ma lies within the 4,480—4,440 age peak for shocked chondrite U-Pb phosphate ages highlighted by previous studies [25, 9]. Primitive meteorites display noticeably reset upper intercept phosphate UPb ages, clustering strongly at around 4,480—4,440 Ma (Figure 6) [10]. The 4,480—4,440 Ma age cluster consists of 10 meteorites (out of 12 with published SIMS ages - Supplementary File 1). The 4,480—4,440 Ma cluster is also diverse, comprising meteorites from at least 4 asteroidal parent bodies (brachinite, carbonaceous, LL, and L ordinary), ruling out simple repeat sampling of an event affecting a single parent body. Of the 6 highly shocked meteorites, 5 plot within the 4,480—4,440 Ma age cluster. This age cluster also notably contains several weakly shocked meteorites (Figure 4b). However, several lines of evidence nonetheless link all of these ages to impact-induced metamorphism.

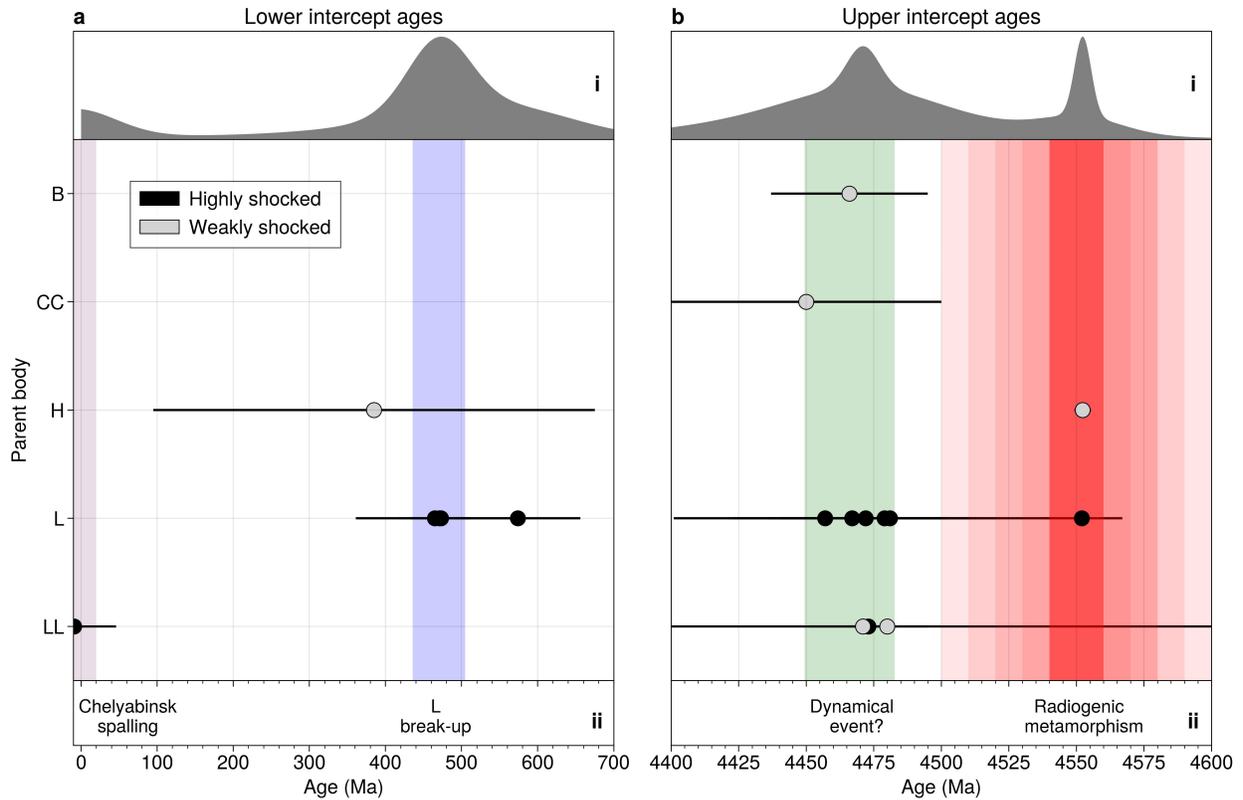

**Figure 6 Compiled shock and phosphate U-Pb age data for meteorites** in a) recent Solar System history and b) early Solar System history. a,b-i) Stacked Gaussian probability distributions of meteorite phosphate U-Pb ages. a,b-ii) Compilation of meteorites and their U-Pb phosphate ages. We divide meteorites by shock stage (see Electronic Appendix) into weakly shocked (S1-3) and highly shocked (S4-6), with the latter being sufficient for full phosphate U-Pb resetting [38]. Estimated limits for temporal range of parent body thermal metamorphism is shaded in red. Lower intercept evidence for an L-type break-up event is found at around 470 Ma, as well as a comparatively recent LL-type break-up event involving Chelyabinsk. A cluster of upper intercept ages is clearly defined at 4,480—4,440 Ma, which may record a major dynamical event in the Solar System. Data from refs: [25, 9, 45, 42, 43, 39, 46, 57, 58, 59]. Error bars are 2$\sigma$.

A plausible mechanism for producing young phosphate U-Pb upper intercept ages in otherwise weakly shocked meteorites is being exposed to a fluid flow. Unequilibrated asteroidal material may be strongly chemically reactive during fluid flow induced by mild impact-induced heating; conditions which are suitable for phosphate nucleation and growth [42]. Thus, apparently reset phosphate U-Pb ages can be produced by new growth, requiring less extensive heating than is needed to fully diffuse Pb from a pre-existing phosphate grain. Such hydrothermal activity can occur in otherwise thermally quiescent asteroids following impact events. Looking to the specific low shock samples that plot around 4,480—4,440 Ma in Figure 6b, both Dar al Gani 978, a carbonaceous chondrite, and Graves Nunataks 06128, an ungrouped achondrite of possible brachinite affinity, preserve evidence of late stage hydrothermal activity that produced phosphates [42, 43].

Conversely, especially prolonged heating may erase (anneal) textural evidence of shock in a meteorite [44]. Dishchii'bikoh, an LL7 chondrite in the 4,480-4,440 Ma cluster, is severely metamorphosed but displays limited shock-related features, such as thin melt veins that cross-cut primary metamorphic features [45]. However, phosphates in Dishchii'bikoh display Pb-Pb and U-Pb ages that are within error of one another (at around 4,480—4,440 Ma). It is therefore likely that phosphates in Dishchii'bikoh either formed or were completely stripped of Pb at 4,470 Ma, corresponding to a significant thermal perturbation of the LL parent asteroid via impact at this time.

The phosphate texture-age record also contains highly shocked meteorites with older upper intercept ages measured for host rock phosphates, compared to younger upper intercept ages for melt-vein entrained grains, e.g., Suizhou, 4,547 ± 19 for host rock phosphates versus 4,481 ± 30 Ma for melt-entrained phosphates [46]. Furthermore, there are examples of weakly shocked meteorites with upper intercept ages consistent with radiogenic cooling, which also preserve lower intercept ages, e.g., Richardton, 4,552.3 ± 3.1 Ma and 385 ± 290 Ma [47]. These pieces of evidence strongly indicate that, as in Chelyabinsk, reset upper intercept phosphate U-Pb ages in primitive meteorites track the intense post-shock heating associated with major impacts, whereas lower intercept ages do not require such conditions to be produced [32, 33, 36]. Finally, we find that all reported upper and lower intercept phosphate U-Pb ages for chondritic meteorites cluster in ancient and recent Solar System history.

There are presently no examples of upper or lower intercept phosphate U-Pb ages that lie between 3 and 1 Ga reported for primitive meteorites. Given the conditions we interpret to have produced upper versus lower interpret ages, we conclude that (1) the abundance of primitive asteroidal material with fully reset phosphates, and thus the frequency of large highly energetic impact events, steeply declined after around 4,440 Ma, and that (2) owing to a short residence time of material on Earth-crossing orbits, and the scarcity of fossil meteorites on Earth, there is a strong sampling bias in our collections towards more recently liberated asteroidal material, with young lower intercept ages. The L parent body disruption event is an outlier in this regard, having produced a large amount of material preserved in the fossil meteorite record [16], and which continues to fall to Earth today [9].

*Implications for dating ancient and recent asteroid collisions*

We have presented evidence that Chelyabinsk phosphate texture-age relationships robustly record an early energetic collision and a recent spalling event. Our interpretation of a recent spalling event involving the LL chondrite body - as revealed by Chelyabinsk phosphate U-Pb lower intercept ages and fracture-associated patchy CL textures - is also supported by evidence from Chelyabinsk Ar-Ar systematics [30] and cosmic ray exposure ages [28]. Geologically recent interaction between asteroidal parent bodies is well supported by observations of the present day asteroid belt, which suggest numerous recent (less than 50 Ma) collisions involving chondritic material [48, 49, 50]. Phosphate U-Pb lower intercept ages may therefore date events of some significance in recent inner Solar System collisional history.

Whilst not microtexturally constrained, the extensive dating work performed by Yin et al

[9] on the Novato L6 chondrite reveals a potentially robust lower intercept phosphate U-Pb age, defined by data lying close to the concordia, that is within error of the Ar-Ar and fossil meteorite age peak observed for meteorites from this parent body [23]. However, Ar-Ar methods both for Chelyabinsk and the wider meteorite record also return numerous ages that are not clearly evidenced in either the phosphate U-Pb system or by mineral textures (Supplementary Figure 17-18). It can therefore be argued that phosphate minerals offer an archive of ancient and recent thermal events that may be absent or overprinted in the Ar-Ar system. However, we cannot yet place equivalent levels of confidence in upper versus lower intercept phosphate U-Pb ages.

Our results reveal that interpreting the detailed structure of the meteoritic phosphate lower intercept U-Pb age record will require the use of microtextural constraints to sub-divide isotopic data for regression, e.g., the significantly revised lower intercept age obtained in this study in comparison to former studies of Chelyabinsk phosphates [26]. Currently published lower intercept ages that lack microtextural context must therefore be treated with some caution, especially if obtained after regressing weakly discordant data. Conversely, similar phosphate U-Pb upper intercept ages are obtained during regression regardless of how isotopic data is sub-divided (Figure 4). Whilst microtextural context is needed to correctly interpret the geological histories of individual meteoritic phosphates, our approach reveals that reset phosphate upper intercept U-Pb ages are a robust archive of ancient energetic collisions. Phosphate analyses therefore support that primitive asteroids experienced a pulse of high energy collisions between 4,480—4,440 Ma (Figure 6b), which may indicate Solar System reorganisation at this time, e.g., Earth-Moon formation, or the migration of giant planets [9, 10, 11].

Overall, our results bolster the use of phosphate texture-age relationships in deciphering asteroidal collision histories, where a relative sequence of impacts can be established using mineral textures and put into chronological context using spatially-resolved dating techniques. Phosphate U-Pb lower intercept ages of fracture-damaged phosphate domains emerge from our work as a valuable tool to probe recent disruption events. Here, further textural constraints on structure-chemistry relationships in damaged versus pristine phosphate grains will be vital for understanding mechanisms of phosphate Pb-loss. By comparison, phosphate U-Pb upper intercept ages in highly shocked meteorites are a remarkably robust archive of ancient and intensive collisional reheating. Determining the origin and significance of clustered upper-intercept ages will require leveraging improved phosphate U-Pb age statistics, constraints on phosphate response to shock and post-shock metamorphism, and dynamical simulations coupled to models of diffusion-driven phosphate U-Pb age resetting behaviour [7, 8, 38, 51]. In resolving a collision history for the Chelyabinsk meteorite, we demonstrate the importance of linking textural analysis with mineral age data when tracing collisions via the meteorite record. In future, combined phosphate texture-age analysis has the potential to access and interpret detailed asteroidal chronologies of both ancient and recent Solar System evolution.

Methods

U-Pb phosphate SIMS analyses

U-Pb dating of apatite was carried out using the CAMECA IMS 1280 at the Institute of Geology and Geophysics, Chinese Academy of Sciences (IGGCAS). The $O^-_2$ primary ion beam was accelerated at -13.8 kV with a current of $10^{-12}$ nA. The Gaussian illumination mode was used in order to evenly sputter material over the analytical area. The spot diameter was 10×15 $\mu m^2$. Positive secondary ions were extracted with a 10 kV potential. Four magnetic field sequences were used to collect secondary ions $_{40}Ca_{31_2}\,P_{16}O^+_3$, $_{204}Pb^+$, $_{206}Pb^+$, $_{207}Pb^+$, $_{238}U^+$, $_{232}Th_{16}O^+$, $_{238}U_{16}O^+$ and $_{238}U^{16}O^+_2$ [52]. The $^{40}Ca^{31}_2\,P^{16}O^+_3$ peak was used as a reference peak for centering the secondary ion beam as well as for making energy and mass adjustments. NW-1 apatite standard (1160 ± 5 Ma) was used for U-Pb fractionation calibration [52] and ages were calculated using IsoplotR [52, 53]. Further details can be found in [54]. We identified fracture-damaged versus pristine phosphate domains using SEM BSE + CL images (see Supplementary Information Figures 1-9).

Compilation of meteorite shock age/stage data

Textural evidence of shock is often classified using the shock stage scheme [55, 56], and provides some context for age data obtained using a given sample. We performed a comprehensive survey of meteorite shock ages and shock stages. We report age uncertainties to 2 $\sigma$. Where shock stages were not directly reported in the meteorite, or where conflicting shock stages were assigned, we applied a simple set of classification rules: 1) petrologic type 7 and impact-melt samples are assigned shock stage 6 (highest possible), 2) more recent classifications take precedence, 3) samples with some noted presence of shock darkening and melt veins are cautiously assigned shock stage 3 (moderate shock), 4) samples with extensive but not complete development of shock melt (i.e., S46 samples, such as Chelyabinsk) are given a single shock stage classification of 5. We then group meteorites into highly shocked (S4-6) and weakly shocked (S1-3), which corresponds to the conditions above and below the threshold for phosphate U-Pb resetting determined by Blackburn et al (2017) [38]. This approach simplifies visual presentation of the shocked meteorite record. We note that the formal guidance for assignment and interpretation of meteorite shock stages has varied over time [55, 18]. However, our compilation broadly includes recently studied meteorites, for which shock stages assignment may differ only subtly in the literature (e.g., reference to Chelyabinsk as an S4-6, or S5, meteorite). Thus, the essential features of the record are robust in the face of minor disagreements on meteorite shock stage assignments in the literature.

1111/maps.12193.

[52] Qiu-Li Li et al. "In-situ SIMS U–Pb dating of phanerozoic apatite with low U and high common Pb". In: *Gondwana Research* 21.4 (2012), pp. 745–756. ISSN: 1342-937X. DOI: 10.1016/j.gr.2011.07.008.

[53] Pieter Vermeesch. "IsoplotR: a free and open toolbox for geochronology". In: *Geoscience Frontiers* 9.5 (2018), pp. 1479–1493. ISSN: 1674-9871. DOI: 10.1016/j.gsf.2018.04.001.

[54] Qin Zhou et al. "Geochronology of the Martian meteorite Zagami revealed by U–Pb ion probe dating of accessory minerals". In: *Earth and Planetary Science Letters* 374 (2013), pp. 156–163. ISSN: 0012-821X. DOI: 10.1016/j.epsl.2013.05.035.

[55] Dieter Stöffler, Klaus Keil, and Scott Edward R.D. "Shock metamorphism of ordinary chondrites". In: *Geochimica et Cosmochimica Acta* 55.12 (1991), pp. 3845–3867. ISSN: 00167037. DOI: 10.1016/0016-7037(91)90078-j.

[56] Jörg Fritz, Ansgar Greshake, and Vera A. Fernandes. "Revising the shock classification of meteorites". In: *Meteoritics & Planetary Science* 52.6 (2017), pp. 1216–1232. ISSN: 1945-5100. DOI: 10.1111/maps.12845.

[57] Yunhua Wu and Weibiao Hsu. "Petrogenesis and In Situ U-Pb Geochronology of a Strongly Shocked L-Melt Rock Northwest Africa 11042". In: *Journal of Geophysical Research: Planets* (2018). ISSN: 2169-9097. DOI: 10.1029/2018je005743.
34

Acknowledgements  C. W. acknowledges NERC and UKRI for support through a NERC DTP studentship, grant number NE/L002507/1. S. H. acknowledges support from the National Natural Science Foundation of China (grant number 41973062) and the key research program of the Institute of Geology and Geophysics, CAS (IGGCAS-201905). A. S. P. R. acknowledges support from Trinity College Cambridge. M. A. acknowledges funding from the UK Science and Technology Facilities Council (STFC) grants #ST/P000657/1 and #ST/T000228/1. Dr Iris Buisman and Dr Giulio Lampronti are thanked for their assistance with microscopy work.

Competing Interests  The authors declare that they have no competing financial interests.

Correspondence  Correspondence and requests for materials should be addressed to C.R.W (email: crw59@cam.ac.uk).